\documentstyle[prl,preprint,aps]{revtex}
\tightenlines
\begin{document}
\draft
\title{Self organization in a minority game: the r\^ole of memory and a
probabilistic approach}
\author{E. Burgos\cite{byline} and Horacio Ceva\cite{byline2}}
\address{Departamento de F\'{\i}sica, Comisi\'on Nacional de
Energ\'{\i}a At\'omica,\\
Avda. del Libertador 8250, 1429 Buenos Aires, Argentina}
\date{9 March, 2000}
\maketitle

\begin{abstract}
A minority game whose strategies are given by probabilities $p$, is replaced
by a `simplified' version that makes no use of memories at all. Numerical
results show that the corresponding distribution functions are
indistinguishable. A related approach, using a random walk formulation,
allows us to identify the origin of correlations and self organization in
the model, and to understand their disappearance for a different strategy's
update rule, as pointed out in a previous work.

\end{abstract}


The minority game (MG), introduced by Challet and Zhang \cite{Ch+Z},
addresses the problem of self organization of a population {\it without }%
direct interactions between its members, but with a feedback mechanism
related with its {\it collective}\ behavior. Each person (dubbed \ `agent',
because of the use of the model in economic problems) has to choose from a
simple alternative, without knowing what the other agents will do. At the
end of the game, there are two groups, one for each alternative:\ agents
belonging to the smaller group (the \ `minority') will be the winners.
Feedback is established by a reward system for winners and losers.

Different methods to choose one or the other alternative give rise to
different versions of the model. It is usual to refer to these methods as\
`strategies'. In the following, we will use the model proposed by Johnson
{\it et.al.} \cite{Johnson}. In this version, each agent knows beforehand
the previous $m$ outcomes (a `history'), of the game, as well as the `next
move' of the most recent occurences of all $2^m$ possibilities. What is
distinctive of Johnson's {\it et.al.} formulation is the assignement to each
of the $N$ agents of a single number as its strategy, $p_j$ $(0\leq p_j\leq
1)$: given a history, each agent will either choose the same outcome as that
stored in the memory, with probability $p_j$, or will choose the opposite
with probability $q_j=1-p_j$ $(1\leq j\leq N).$

Winners, {\it i.e.} those in the minority group, will gain $G$ points ($%
G\equiv 1$ in Ref.\cite{Johnson} ); those in the majority group lose a
point. Strategies can be modified, following the evolution of the game:\ if
the number of points of an agent is below a threshold value $d\leq 0$, his \
`account' is reset to zero, and he/she gets a new strategy, whose value $%
p^{\prime }$ is chosen with an uniform probability from the interval $%
(p-R/2,p+R/2)$, where $|R|\leq 2;$ in what follows we will use the simpler
notation $p\rightarrow p^{\prime }=p\pm \Delta p.$ Whenever necessary, we
used reflective boundary conditions.

Johnson {\it et.al.} have shown that, as a result of correlations, the
system self organizes mainly in two distinct groups of agents, with extreme
values of their strategies:\ $p\approx 0$ or $p\approx 1.$ The frequency
distribution $P(p)$ is shown in Fig.\ref{fig1}\cite{scatter}. Moreover, they
also found that their results did not change if memories are not updated, or
even if they are randomly chosen; a similar result was obtained by Cavagna
\cite{Cavagna} in relation with the work of Challet and Zhang\cite{Ch+Z}.

In this work, we show that a \ `simplified' version of the model, making no
use of the memories, also displays self organization, and the resulting
distribution $P(p)$ is indistinguishable from the original model, for all
odd $N\geq 3$. Another approach (called `probabilistic' in what follows)
serves to present a rather detailed interpretation of this MG, explaining
the mechanisms establishing correlations between the agents, and their
relations with the rules of the game. This also allows us to explain the
smallness of correlations found in a previous publication, where a different
update rule of the strategies was used \cite{hc}.

In our \ `simplified' version of the model, agent $j$ chooses one of the two
options (option ``1'', say) with probability $p_j,$ or the other (option
``0'') with probability $q_j=1-p_j$,{\it \ without making recourse to any
history at all}. All other rules, like the determination of the minority,
the reward system, the upgrade of strategies $p$, $etc.$ are the same as
before. In Fig. \ref{fig1} \ we compare results obtained with this version
and with the original formulation, for $N=101$, $d=-4$, $R=0.2$ . A single
realization of the game involved $n_t=10^5$ time steps, and the distribution
was averaged over $n_s=10^4$ samples. As it was already mentioned, both
results are indistinguishable.

Even if one lets the memories totally outside of the game, there still are
many parameters in the model (the variables $G,d,N,R)$, and it is worth to
see how it depends on these variables. All our simulations were made in such
a way that we can assure that the expected numerical fluctuations for the
initial, uniform distribution, were small. To this end, we requested that
the standard deviation $1/\sqrt{Nn_s/c}$ be no greater than 0.02 . We always
used c=100 channels in $p$ (each of width 0.01). Moreover, to be able to
compare results for different values of $N$, we have normalized all our data
so that $\int P(p)dp=1.$

We collected data for $N=3,5,7,9,11,21,51$ and $101$, $d=0,-2,-4$ and $-5$,
for fixed values of $R=0.2$ , and $G=1$. Figure \ref{fig2} shows the density
values for both extrema, $P(0),P(1)$ $vs$ $1/N$. It is rather clear that $%
N=101$ is already near the \ `thermodynamic' limit $(N\rightarrow \infty )$.
Also, there is only a moderated dependence on the value of the threshold $d.$

Results for the `simplified' version of the model show that we need not to
consider memories, and constitute the starting point for the following
development.

Our second, probabilistic, approach originated in the observation that the
change of strategies, $p\rightarrow p^{\prime },$ can be thought of as the
movement of points in $p$ space, suggesting the formulation of the model as
some kind of generalized random walk (RW). It is of some interest to mention
that there is a formal similarity of $P(p)$ with a property of a RW;
specifically, this is the case of the expresion for the probability of the
last visit to the origin, $x=0,$ for such a system \cite{Feller}. In the
simplest version of a RW, a point moves regularly with a constant step $S$,
randomness being only related with the sign of $S.$ Applied to the case of $%
N $ points randomly distributed in a one dimensional (1D) box, it produces a
uniform distribution in space, $P(x)=P_0$. In a more general case, the
probability $m(x)$ to move a point will be, in general, a function of
position $x$. This would be the case, for instance, of a system with
absorbing walls, or a gas with a thermal gradient. Moreover, there are
situations where the movement needs not to be regular (in time). The
stationary state would be established when $m(x)P(x)=constant$.

We will consider the movement of $N$ agents in a 1D space of probabilities, $%
p,$ with a variable step $\Delta p$, as it was already made in Johnson's
formulation of the model. On the other hand, we will use the set $%
\{p_i,G,d\} $ to decide if an agent moves or not. It is possible to write $%
\mu _i,$ the probability of agent $i$ to be in the majority, in terms of $%
\{p_j\}$. If $G=d=0$, then it is simple to see that $\mu _i$ is equal to the
probability $m(p_i)$ to move the agent, $i.e.$ $p_i\rightarrow p_i^{\prime
}. $ In the following we will refer to $m$ as the {\it mobility}. Note that,
in general, the actual value of $m$ will depend on all the $p_j;$ we choose
to write $m(p_i)$ for agent $i$, to emphasize that its value changes with
the position of the agent in $p$-space. We cannot find a closed expresion
for the mobility if $G$ and $d$ are $\neq 0$; nevertheless, as the mobility
must follow in general the behavior of $\mu $, we still can use it to
describe the system's behavior.

Let us consider the case $N=3$, for simplicity. The system is characterized
by the strategies $p_{1},p_{2},p_{3}$ (and $q_{i}=1-p_{i}$). The
probabilities for every agent to be in the minority are

\begin{equation}
\lambda _{1}=p_{1}q_{2}q_{3}+q_{1}p_{2}p_{3},\ \ \lambda
_{2}=q_{1}p_{2}q_{3}+p_{1}q_{2}p_{3},\ \lambda
_{3}=q_{1}q_{2}p_{3}+p_{1}p_{2}q_{3}  \label{prob1}
\end{equation}

and the corresponding probabilities of being in the majority are $\mu
_{i}=1-\lambda _{i}$.

These expressions can be easily generalized to the case of $N=2n+1$ agents.

\bigskip Introducing $x_i=p_i/q_i$, $y_i=1/x_i$, $U=\prod_{j=1}^{j=N}p_i$, $%
Q=\prod_{j=1}^{j=N}q_i$ it is \cite{zeros}

\begin{eqnarray}
\lambda _i &=&x_iQ(1+\sum_{k_2}x_{k_2}+\sum_{k_2,k_3}x_{k_2}x_{k_3}+...+
\sum_{k_2,k_3,..,k_n}x_{k_2}x_{k_3}...x_{k_n})  \nonumber \\
&&+y_iU(1+\sum_{k_2}y_{k_2}+\sum_{k_2,k_3}y_{k_2}y_{k_3}+...+
\sum_{k_2,k_3,..,k_n}y_{k_2}y_{k_3}...y_{k_n})  \label{prob2}
\end{eqnarray}

Equations(\ref{prob1})-(\ref{prob2}) show that $\mu _i$ depends on all the
strategies, $\{p_j\}$. In other words, it illustrates that the origin of
correlations in the distribution $P(p)$ can be traced back to the rules
defining the minority game.

We have made numerical simulations based on Eqs.(\ref{prob1})-(\ref{prob2}),
for $N=3,5,7,9$ and $11$: at every step \ of the game, an agent gains
(loses) one point with probability $\lambda $ $(\mu )$. This procedure makes
explicit {\it how }agents relate their behaviors through the strategies.
Figure \ref{fig3} shows results for $N=$ $11$, together with the
corresponding results from our `simplified' version. The similarity between
the results obtained with both methods it is remarkable; the small
differences seen in the very narrow region near both extrema $(p\approx
0,p\approx 1),$ are probably due to the noise attributable to our
simulations. This is not to say that this approach is identical to the
original model. In fact, we can only expect the probabilistic approach to be
equivalent to the `simplified' version {\it in a statistical sense}, but it
is not possible to compare both methods at each time step. To illustrate
this difference, notice that in this formulation one will accept some
outcomes which are not allowed in the original MG; thus, for instance, as
agents win a point with probability $\lambda _i,$ there exists a finite
probability, $W_v(N)\neq 0,$ that the majority of the agents can win a
point, in an {\it apparent} violation of the basic rule of the game (indeed,
it can even happen that {\it all} agents are simultaneously winners). If $%
N=3,$

\begin{equation}
W_v(3)=\lambda _1\lambda _2\lambda _3+\mu _1\lambda _2\lambda _3+\lambda
_1\mu _2\lambda _3+\lambda _1\lambda _2\mu _3  \label{prob3}
\end{equation}

and similar relations for all $N.$

Using Eq.(\ref{prob2}), and the generalization of Eq.(\ref{prob3}), we
calculate for a uniform distribution that $W_v(3)\approx 0.14$, while it can
be estimated that $W_v(\infty )\leq 0.25$. Similar values are obtained for
non uniform distributions with a shape analogous to that of Fig.\ref{fig3}.

Figure \ref{fig4} has results for $N=3,$ $d=0,-1$ and $G=0,1.$ In this case,
results obtained with the `simplified' version (not shown here) are
indistinguishable from those coming from Eq.(\ref{prob1}). If $G=0,$ the
ensuing self-organization is small;\ we have verified the same type of
behavior for all $N\leq 11$\cite{precision}. On the contrary, if $G=1$
self-organization is very important. In both cases, $d$ has a smaller
influence on the behavior of the system. We have included results for $%
N=3,d=G=0$ in Fig. \ref{fig4}, because in this case $m(p)=\mu (p),$ and it
is possible to get a clear picture of the resulting (small) organization.
Assume agents are numbered so that $p_1<p_2<p_3$, and consider the case $%
p_1<1/2,$ $p_3>1/2$. It follows from Eq.(\ref{prob1}) that $m_2>m_1,m_3$. \
This describes a situation \ where agents near $p\approx 0$ and $p\approx 1$
have a tendency to remain in their positions, while the agent in between
moves more frequently. Eventually this will change if, as a result of the
movement, either $p_2<p_1$ or $p_2>p_3,$ increasing the accumulation of
agents near $p=0$ and $p=1$ \cite{p<1/2}. Incidentally, we can use this
picture to understand why the use of a different updating rule can destroy
self-organization, as previously reported by one of us \cite{hc}. Assume the
same situation as before, $i.e.$ an agent with $p_2\approx 1/2$ and high
mobility, and two agents near $p_1\approx 0$ and $p_3\approx 1$, with
smaller mobility. If now strategies are updated using $p^{\prime }=1-p\pm
\Delta p,$ as the agent with $p_1\approx 0$ moves, he will go near $%
p_1\approx 1.$ It is easy to see from Eq.(\ref{prob1}) that the
corresponding mobilities are $m_1\approx m_3\approx 1>m_2.$ In words,
correlations are broken in a single step, so that only a very tiny
indication of self-organization remains. It is worthwhile to mention that
this type of evolution can no longer be sensibly described as a random walk.

The probabilistic approach provides a natural starting point for an analytic
formulation of this model. A sample is described by a point in an $N$
dimensional space of probabilities $p_j$; the set of all the $n_s$ points
can be thougth of as a non interacting gas that evolves towards a stationary
distribution, driven by the rules of the game. We are presently working on
the implementation of these ideas \cite{Pancho}.

In summary, we have made mainly two contributions to the knowledge of this
version of the MG: $(i)$ we have shown the irrelevance of memory in the
resulting self organization of the system; in this respect, therefore, this
version of the MG behaves differently than that of \cite{Ch+Z}, where it has
been claimed \cite{Cavagna} that all agents {\it need to receive the same
information}, whether it be true or false, to be able to self organize; $%
(ii) $ our probabilistic formulation proved to be a very good approximation
to the model and, equally important, allows us to understand in detail how
the game's rules establish correlations between the agents.

We thank F. Parisi for useful discussions and comments. E.B. was partially
supported by CONICET of Argentina, PICT-PMT0051; H.C. was partially
supported by EC Grant ARG/B7-3011/94/27, Contract 931005 AR.

\begin{figure}[tbp]
\caption{Distribution of strategies $P(p)$ for the original model and for
our `simplified' approximation. $N=101$, $G=1$, $d= -4$, $R=0.2$, $n_t=10^5$
, $n_s=10^4$. In the original model, $m=3$}
\label{fig1}
\end{figure}

\begin{figure}[tbp]
\caption{Extreme strategies $P(0)$ and $P(1)$ vs $1/N$. Filled symbols refer
to $P(0)$, open ones to $P(1)$}
\label{fig2}
\end{figure}

\begin{figure}[tbp]
\caption{Distribution of strategies $P(p)$ for our two approximations. $%
N=11, G=1,d= -1, R=0.1, n_t=10^5, n_s=10^4$}
\label{fig3}
\end{figure}

\begin{figure}[tbp]
\caption{Distribution of strategies $P(p)$, in the probabilistic
approximation, for $N=3$. Open symbols refer to $d= -1$, filled symbols to $%
d=0$. The inset shows an enlarged picture for $G=0$}
\label{fig4}
\end{figure}


\begin{references}
\bibitem{scatter}  We have chosen to display our data as discrete points,
because in this form it is possible to appreciate the small differences
between all the methods used here.
\bibitem{zeros}  Indeterminations for $p=0$ or $1$ are only apparent, and
are vey easily eliminated.
\bibitem{precision}  $N\approx 11$ is near the limit allowed to numerical
calculations, both because of the amount of computer time involved and, more
importantly, because of precision considerations
\bibitem{p<1/2}  It can be seen that the mean value of the initial mobility
for $N=3$ is $3/4$; in general, $2/3<(m_1+m_2+m_3)/3<1$. If all $p^{\prime }s
$ are smaller (greater) than $1/2$, a similar analysis shows that all agents
will try to be near $p\approx 1/2$, with a mobility $m\simeq 1-p+p^2,$ which
is even greater than the initial value. This suggest that this configuration
should be unstable.
\end{references}
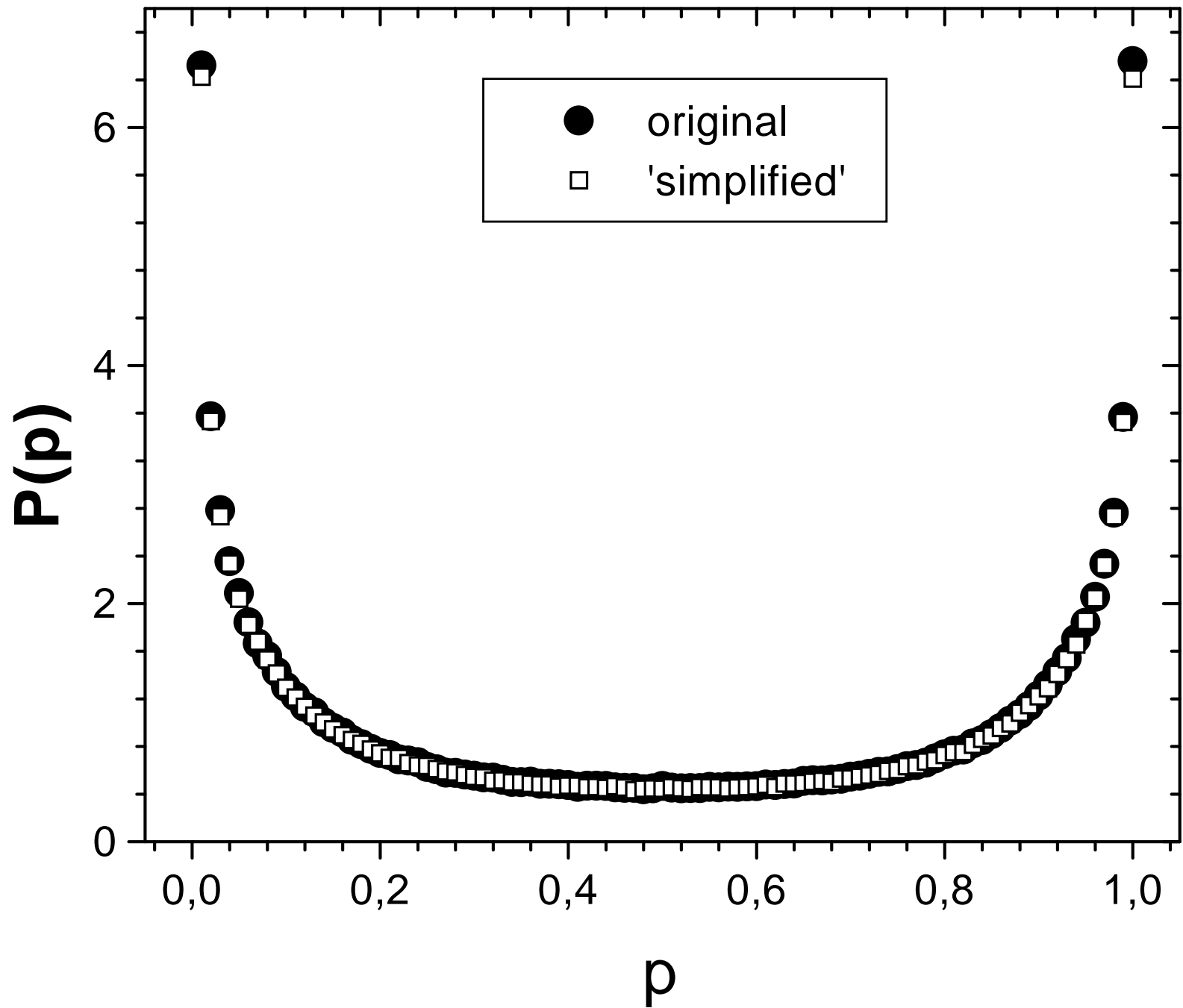
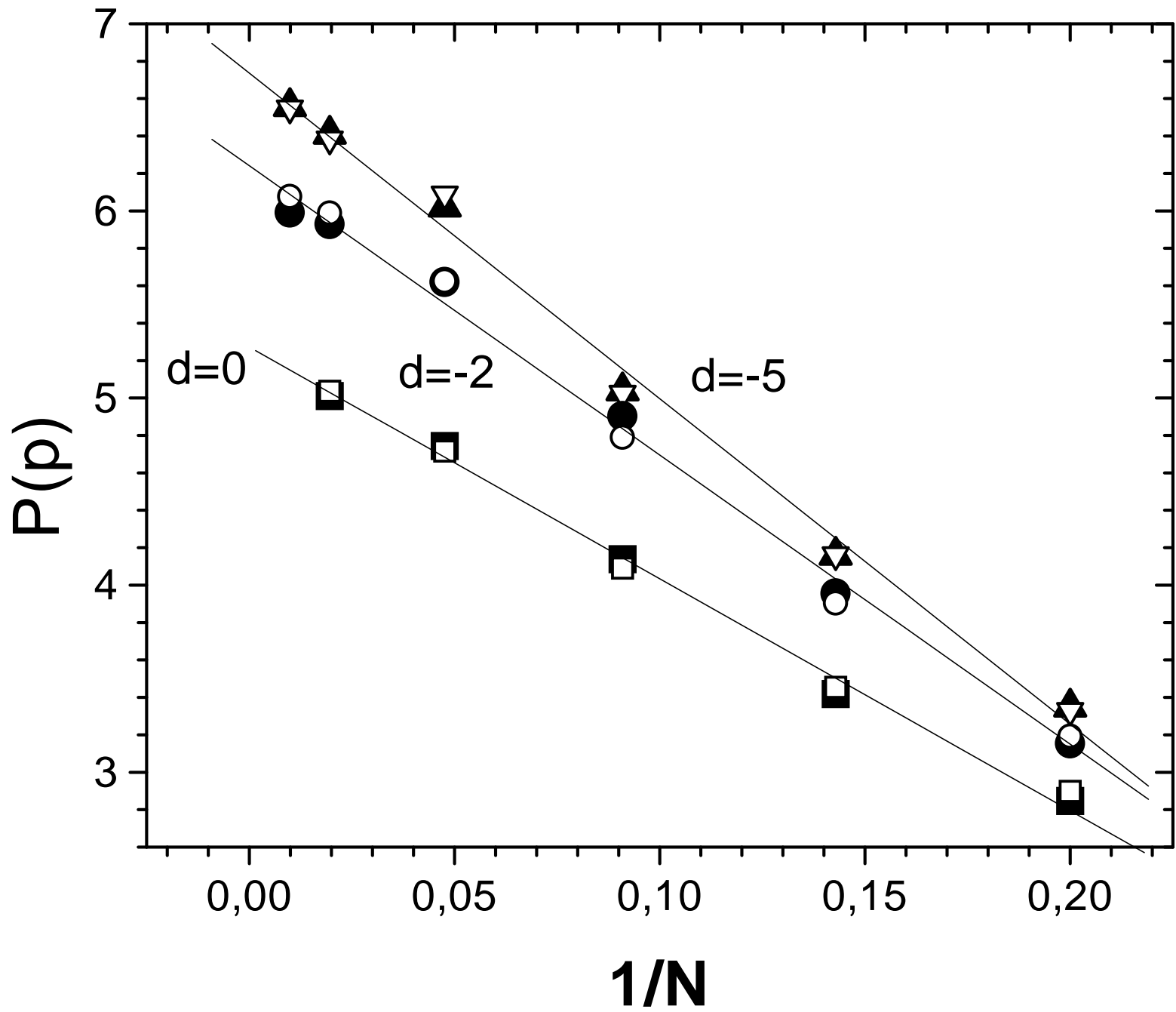
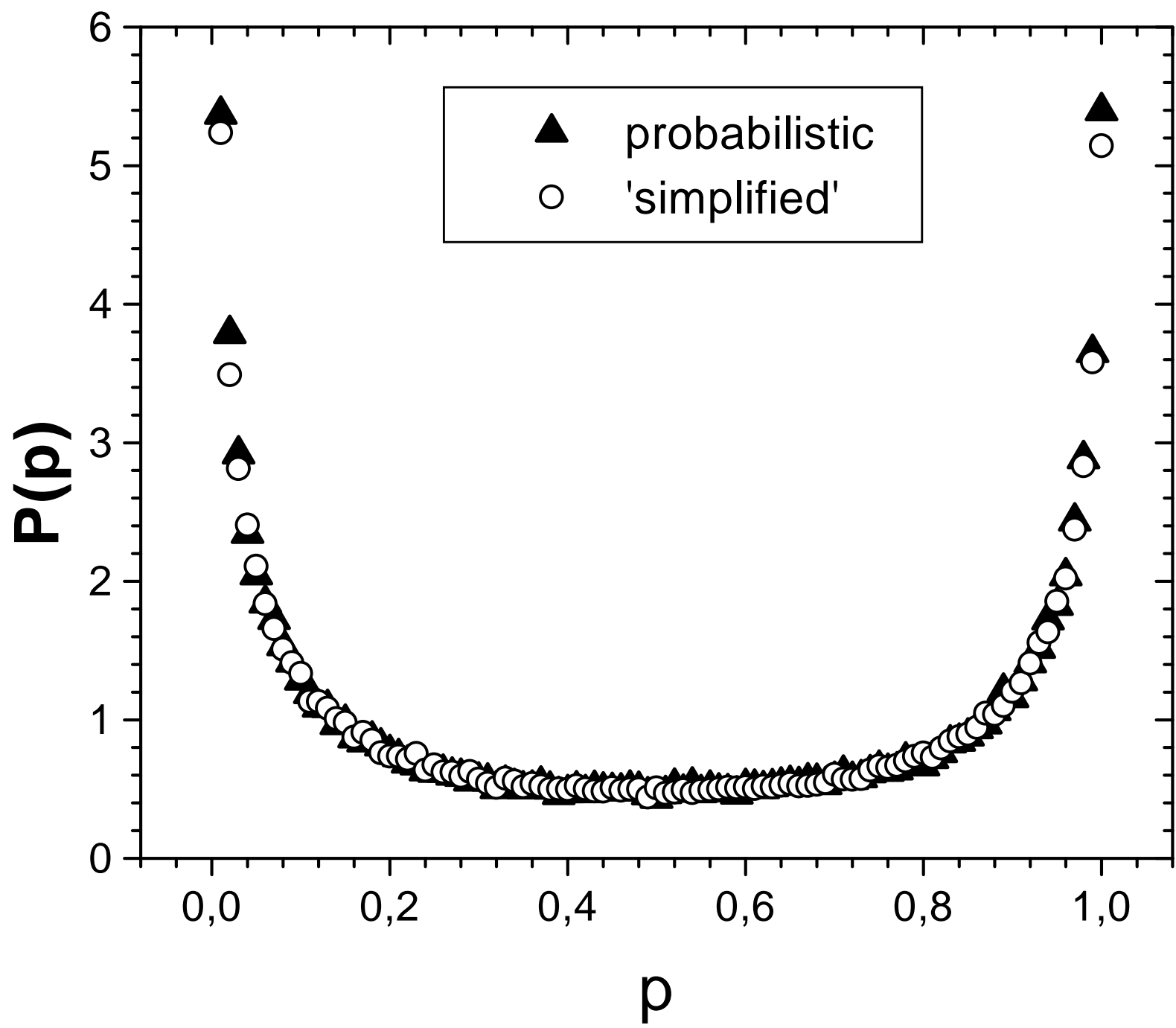
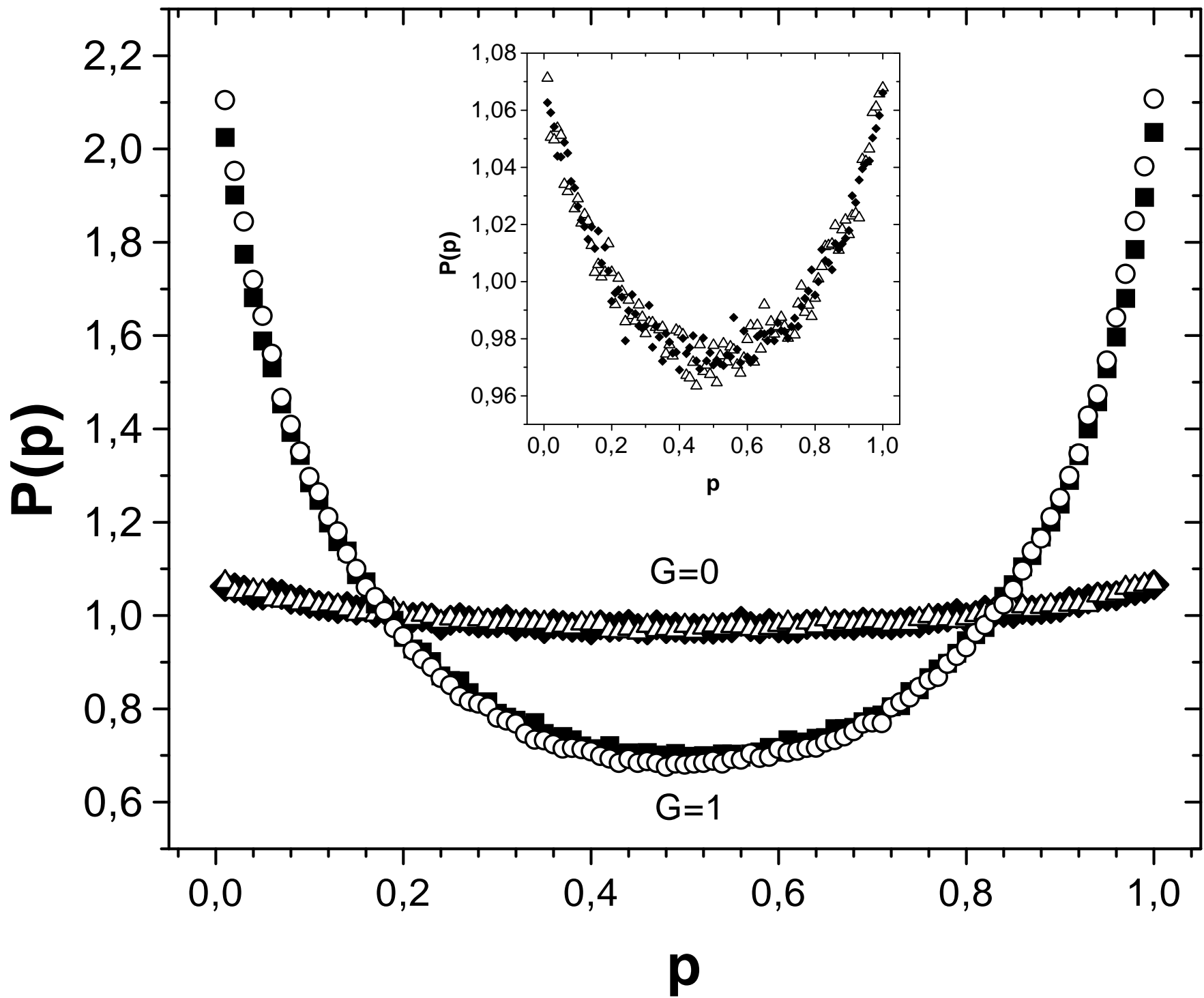
\end{document}